\RequirePackage{fix-cm}
\documentclass[smallextended]{svjour3}       
\smartqed  
\usepackage[hyphens]{url}
\usepackage{natbib}
\usepackage{amsmath,amssymb,amsfonts}
\usepackage{algorithmic}
\usepackage{graphicx, subfigure}
\usepackage{textcomp}
\usepackage{xcolor}
\usepackage[skins,breakable]{tcolorbox}
\usepackage{booktabs}
\usepackage[htt]{hyphenat}
\usepackage[hidelinks]{hyperref}
\usepackage{float}
\usepackage{caption}
\usepackage{threeparttable}
\usepackage[misc]{ifsym}
\usepackage{todonotes}
\usepackage{listings}
\usepackage[normalem]{ulem}

\def\BibTeX{{\rm B\kern-.05em{\sc i\kern-.025em b}\kern-.08em
    T\kern-.1667em\lower.7ex\hbox{E}\kern-.125emX}}
    
\graphicspath{{figs/}}

\newcommand{\fix}[1]{\textcolor{black}{#1}}
\newcommand{\minor}[1]{\textcolor{black}{#1}}

\journalname{Empirical Software Engineering}
\begin{document}

\title{18 Million Links in Commit Messages: Purpose, Evolution, and Decay
}


\author{Tao Xiao \Letter \and Sebastian Baltes \and Hideaki Hata \and Christoph Treude \and Raula Gaikovina Kula \and Takashi Ishio \and Kenichi Matsumoto
}


\institute{
    \Letter~Corresponding author - Tao Xiao
    \at Nara Institute of Science and Technology, Japan\\
    \email{tao.xiao.ts2@is.naist.jp}
    \and 
    Sebastian Baltes
    \at University of Adelaide, Australia\\
    \email{sebastian.baltes@adelaide.edu.au}
    \and 
    Hideaki Hata
    \at Shinshu University, Japan\\
    \email{hata@shinshu-u.ac.jp}
    \and
    Christoph Treude \at
    University of Melbourne, Australia\\
    \email{christoph.treude@unimelb.edu.au}
    \and
    Raula Gaikovina Kula, Takashi Ishio, Kenichi Matsumoto
    \at Nara Institute of Science and Technology, Japan\\
    \email{\{raula-k,ishio,matumoto\}@is.naist.jp}
}

\date{Author pre-print copy. The final publication is available at Springer via:\\
https://link.springer.com/article/10.1007/s10664-023-10325-8}

\maketitle

\begin{abstract}
Commit messages contain diverse and valuable types of knowledge in all aspects of software maintenance and evolution.
Links are an example of such knowledge. Previous work on ``9.6 million links in source code comments'' showed that links are prone to decay, become outdated, and lack bidirectional traceability.
We conducted a large-scale study of 18,201,165 links from commits in 23,110 GitHub repositories to investigate whether they suffer the same fate.
Results show that referencing external resources is prevalent and that the most frequent domains other than \texttt{github.com} are the external domains of \texttt{Stack Overflow} and \texttt{Google Code}.
Similarly, links serve as source code context to commit messages, with \minor{inaccessible} links being frequent.
Although repeatedly referencing links is rare (4\%), 14\% of links that are prone to evolve become unavailable over time; e.g., tutorials or articles and software homepages become unavailable over time. 
Furthermore, we find that 70\% of the distinct links suffer from decay; the domains that occur the most frequently are related to \fix{S}ubversion repositories. 
We summarize that links in commits share the same fate as links in code, opening up avenues for future work. 

\keywords{Commit Messages \and Software Documentation \and Link Sharing \and Link Decay}
\end{abstract}

\section{Introduction}
Developers use commit messages as a means to summarize introduced code changes in natural language \citep{mockus2000identifying, buse2010automatically}.
These descriptions can be used to validate changes, locate and triage bug reports, and trace changes to address software maintenance tasks~\citep{girba2005developers, hassan2008road, d2010commit}.
It is also known that commit messages usually contain diverse types of useful knowledge in terms of how they facilitate the understanding of code changes, e.g., issue, feature, and rationale~\citep{mockus2000identifying, fu2015automated, sarwar2020multi}. 

\begin{figure}[t]
  \centering
    \includegraphics[width=\textwidth]{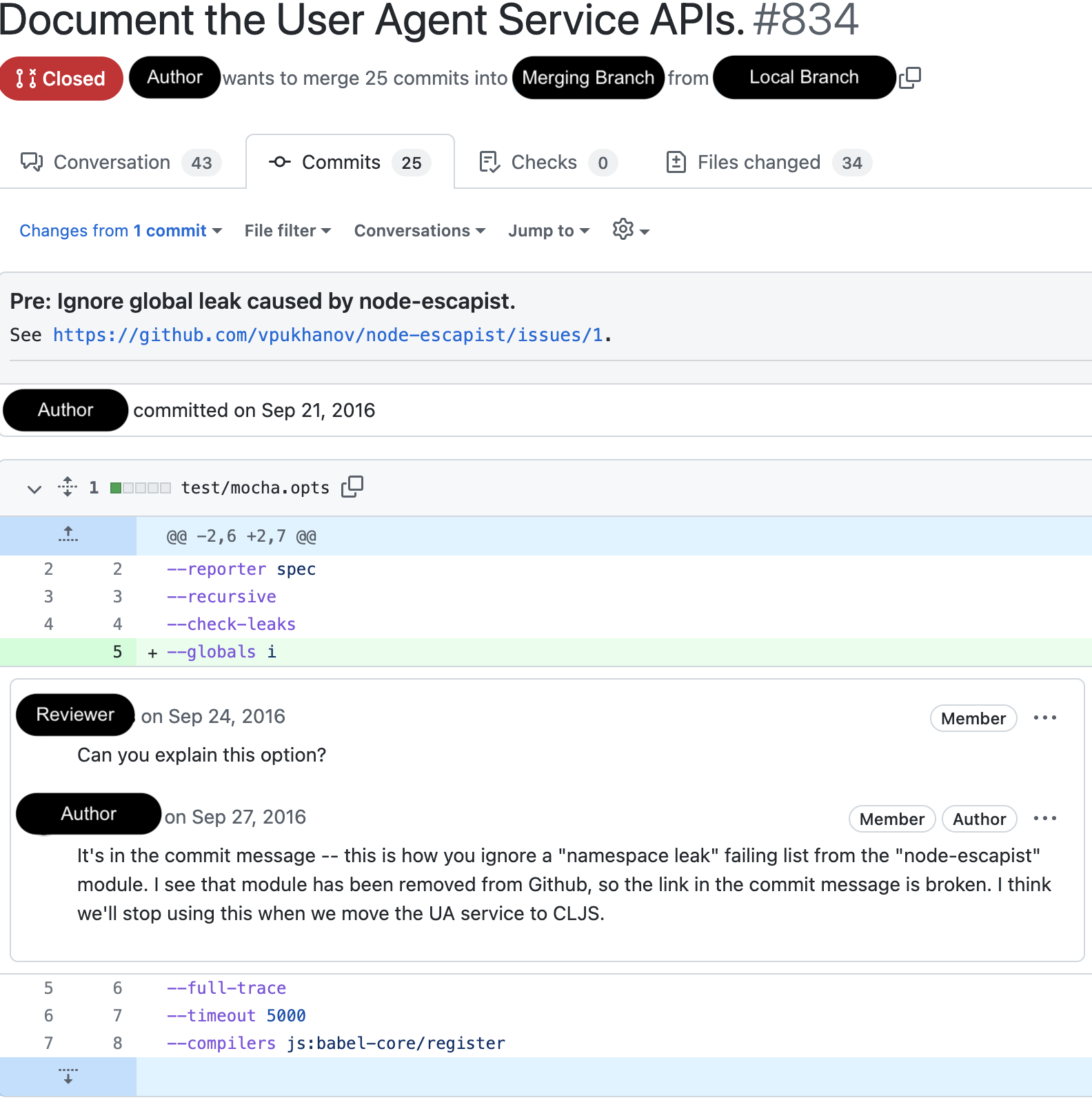}
  \caption{Motivating example of link decay in the commit message.}
  \label{fig:motivating}
\end{figure}

Inspired by the Internet, useful information or knowledge has been represented by links. 
However, the growth of links has brought on more challenges of link decay~\citep{kehoe1998gvu}, digital plagiarism~\citep{barrie2000digital}, and fragile historical \fix{w}eb content~\citep{murphy2007take}.
Recently,~\cite{hata20199} conducted a study to understand the purposes, evolution, and decay of links in source code comments.
They observed decay, insufficient versioning, and lack of bidirectional traceability. 
Link decay is also a common problem in other software artifacts, e.g., Stack Overflow posts~\citep{liu2021broken}.~\cite{liu2022exploratory} also observed that external links are repeatedly referenced in Stack Overflow posts, and they found that repeated external links increase the maintenance effort.
As an important communication channel, commit messages become critical for communicating effectively. Among the knowledge embedded in commit messages, links are special containers that provide additional knowledge for developers in commit messages. Commit messages may play a similar role as source code comments, which communicate information indirectly between code authors and reviewers. Thus, these issues in source code comments may also apply to commit messages. \fix{Figure~\ref{fig:motivating} shows the motivating example of how the link decay in the commit messages causes knowledge loss in the code review process. Such knowledge loss will not only increase review time but also make confusion for future developers.} 

This paper is an extension of~\cite{hata20199}, where we formulate six research questions to establish the understanding of the role of links in commit messages.
We collect 18,201,165 links from commits in 23,110 GitHub repositories. From our quantitative analyses, we find that links are occurring in commit messages, accounting for at least 83\% of GitHub repositories in our study and the internal domain \textsf{github.com} is the most frequently occurring domain in commit messages, followed by \textsf{stackoverflow.com}. Then, we identify the kinds of link targets that are referenced and their purposes served in commit messages through qualitative studies of a stratified sample of 1,145 links. We observe that (i) \minor{inaccessible} links and patch links are the most frequently occurring link target types in commit messages; and (ii) links are often used to serve as source code context or to keep files in sync between versions.
To investigate the phenomenon of links (e.g., repeated link reference, link evolution, and link decay), we conducted a mixed-methods study. Our results show that (i) the behavior of repeatedly referenced links rarely happens in commit messages; (ii) 14\% of the links are prone to evolve to become unavailable over time, e.g., tutorial or article and software homepage are temporarily available software artifacts; and (iii) link decay is a common issue in commit messages. This extension makes the following main contributions:

\begin{itemize}
    \item large-scale and comprehensive studies of around 18 million links to establish the prevalence, link decay, and case study of Stack Overflow link target evolution in commit messages,
    \item a mixed-methods study to identify types of link targets and purposes that links served, links evolution, and reasons why links are repeatedly referenced in commit messages,
    \item a comparison study of the role of links between source code comments and commit messages (i.e., RQs 1, 2, 3, 6, and a case study of Stack Overflow in RQ5), and
    \item a full replication package of our study, including the scripts and data set.\footnote{\url{https://doi.org/10.5281/zenodo.7536500}}
    
\end{itemize}

The rest of the paper is organized as follows. Section~\ref{sec:rq} structures our six research questions and their motivations. Section~\ref{sec:dc} details our data collection process. Section~\ref{sec:md} describes our methods in qualitative and quantitative analyses. Section~\ref{sec:rs} presents our research findings by answering the six aforementioned research questions. Section~\ref{sec:ds} discusses the comparison of the role of links between source code comments and commit messages and recommendations. 
Section~\ref{sec:tv} acknowledges threats to the validity of our study.
Section~\ref{sec:rw} situates
our work in relation to the literature on commit messages, knowledge sharing, and link sharing. 
Section~\ref{sec:cs} draws conclusions and highlights opportunities for future work.




\section{Research Questions}
\label{sec:rq}
In this section, we present our six research questions with the motivation to gain insight on how links are used in commit messages.

\textbf{(RQ1)} \textit{How prevalent are links in commit messages?}

To gain an initial intuition about links and understand the usage of links in commit messages, we set out to quantitatively explore the
distribution, diversity, and spread of these links across different
types of software projects.

\textbf{(RQ2)} \textit{What kinds of link targets are referenced in commit messages?}

\textbf{(RQ3)} \textit{What purpose do links in commit messages serve?}

\textbf{RQ2} and \textbf{RQ3} involve a more qualitative approach to analyze what role links play in what developers use in commit descriptions. Answering these RQs will help characterize why and how developers make references in a commit message.

\textbf{(RQ4)} \textit{To what extent do commit message links get repeatedly referenced?}

\textbf{(RQ5)} \textit{How prone are link targets to evolve over time?}

\textbf{(RQ6)} \textit{How frequent are \minor{inaccessible} links in commit messages?}

\textbf{RQ4}, \textbf{RQ5}, and \textbf{RQ6} investigate the phenomenon of links from the evolutionary and maintenance point of view. We would like to understand whether developers are repeating links, how these links evolve after introducing them, and how many of the links are affected by link decay.

\section{Data Collection}
\label{sec:dc}
To answer the research questions mentioned above, we focus on the same stratified sample of repositories as the previous study used in our earlier study of links in source code files~\citep{hata20199}. \fix{They collected active non-forked repositories for the seven languages from the GHTorrent data set\footnote{MySQL database dump 2019-02-01 from \url{http://ghtorrent.org/downloads.html}.}~\citep{gousios2013ghtorent} using the following
criteria: (i) having more than 500 commits in their entire
history, and
(ii) having at least 100 commits in the most active two
years to remove long-term
less active repositories and short-term projects that have
not been maintained for long.}
The repository list is part of our replication package.

\subsection{Commit Message Collection}
To extract all links from the commit messages in these repositories, we used a bash script to retrieve all commit messages from the \fix{default branch} of these repositories, exporting them to CSV files along with the commit metadata.\footnote{\url{https://github.com/sbaltes/git-log-extractor}} 
Since some repositories were not available because they had been deleted or made private, we obtained \fix{27,263} (93\%) repositories from the given repository list, created in 2018
(see Table~\ref{tab:coll} for details on the different strata).
We then imported those CSV files into Google BigQuery tables, one per programming language. 

\subsection{Link Identification}
Using \fix{the following regular expression} in SQL queries, we extracted \minor{HTTP(S)} URLs\minor{, which is the most common way of hosting or sharing resources,} from collected commit messages stored in Google BigQuery tables:
\begin{lstlisting}[breaklines=true]
(https?:\/\/(?:www\.)?[-a-zA-Z0-9@:%.\+~#=]{1,256}\.[a-zA-Z0-9()]{1,6}\b(?:[-a-zA-Z0-9()@:%\+.~#?&\/\/=]*))
\end{lstlisting}

We identified a total of 18,201,190 links from commit messages, as seen in
Table~\ref{tab:coll}. Since we will conduct a quantitative study on aspects of link domains (RQ1), we exclude 25 false-positive links in this analysis, whose domain is empty \fix{(e.g., ``The learn more link should go to http://...answer=185277.''). These false-positive links are malformed and served as an example of links in commit messages.} As a result, we obtained 18,201,165 links that are used in this study.
The bash and SQL scripts, commit messages, and resulting CSV files are available as part of our replication package. 

\begin{table}[t]
\centering
\caption{Collected repositories and links.}
\label{tab:coll}    
\begin{tabular}{l@{}rr@{}rrr@{}rr}
\toprule

 & \multicolumn{3}{c}{\textbf{\# repositories}} & \multicolumn{3}{c}{\fix{\textbf{\# commits}}} &  \textbf{\# links} \\
 &  \textbf{candidate} &  \multicolumn{2}{c}{\textbf{obtained (\%)}} & \fix{\textbf{All}} & \multicolumn{2}{c}{\fix{\textbf{w/ links  (\%)}}} & \\
\midrule

C           & 2,771 & 2,607 & (94\%) & \fix{122.5M} & \fix{7,256,770} & \fix{ (5.9\%)}  & 8,067,201 \\
C++         & 3,563 & 3,391 & (95\%) & \fix{21.2M}  & \fix{4,188,042} & \fix{(19.7\%)}  & 4,779,742 \\
Java        & 4,995 & 4,701 & (94\%) & \fix{30.4M}  & \fix{1,758,495} & \fix{(5.8\%) }  & 1,891,739 \\
JavaScript  & 7,130 & 6,542 & (92\%) & \fix{13.8M}  & \fix{  634,715} & \fix{(4.6\%) }  &   778,667 \\
Python      & 5,263 & 5,007 & (95\%) & \fix{15.9M}  & \fix{  766,324} & \fix{(4.8\%) }  &   859,396 \\
PHP         & 3,279 & 2,941 & (90\%) & \fix{11.0M}  & \fix{1,335,934} & \fix{~(12.1\%)} & 1,503,529 \\
Ruby        & 2,233 & 2,074 & (93\%) &  \fix{5.9M}  & \fix{  214,647} & \fix{(3.6\%) }  &   320,916 \\
\midrule
\textbf{sum}& \textbf{29,234}& \fix{\textbf{27,263}}& \textbf{(93\%)}& \fix{\textbf{220.8M}} & \fix{\textbf{16,154,927}} & \fix{\textbf{(7.3\%)}} &  \textbf{18,201,190} \\

\bottomrule
\end{tabular}
\end{table}





\section{Method}
\label{sec:md}
In this section, we describe the mixed-methods procedure that includes \fix{quantitative analysis (Section~\ref{sec:qan} for RQs 1, 5, and 6) and qualitative analysis (Section~\ref{sec:qa} for RQs 2, 3, and 4).}

\subsection{Quantitative Analysis}
\label{sec:qan}
To get an intuition of links and their usages in commit messages (\textbf{RQ1}), and investigate the phenomenon of the evolution of link targets (\textbf{RQ5}) and link decay (\textbf{RQ6}), we conduct quantitative analyses of 18,201,165 links and a statistically representative and stratified
sample.

\paragraph{Link Prevalence (RQ1)} For RQ1, we conducted three quantitative studies on aspects of link existence, domain popularity, and popular domains in our data set. For link existence, we calculate the ratio of repositories with at least one link among seven programming languages. For domain popularity, we calculate the distribution of the number of different domains per repository. Median values are used to measure popularity. For popular domains, we calculate the top ten frequent repositories by counting only once in each repository.

\paragraph{Link Target Evolution (RQ5)} To address RQ5, we conducted a quantitative study to investigate the evolution of link targets. We use the Wayback Machine JSON API\footnote{\url{https://archive.org/help/wayback_api.php}}\footnote{\url{https://github.com/sbaltes/wayback-machine-retriever}} to obtain the closest snapshots of links from our 1,145 samples that are used in \fix{qualitative analysis from RQs2--4} in the next two years when we retrieved our data set (e.g., March 16, 2020, March 16, 2021, and March 16, 2022). 
This API will return a JSON object if the given link is archived and currently accessible in the Wayback Machine, and it will return an empty JSON object if the given link is not archived or currently not accessible. Then, we compare these availability statuses to the coding results from RQ2.

\paragraph{Link Decay (RQ6)} For RQ6, we conducted a quantitative study on aspects of link decay in our data set. Out of the obtained 18,201,165 links, there are 6,667,207 distinct links. To investigate the number of \minor{inaccessible} links in commit messages, we accessed all \fix{w}eb content of the 6,667,207 unique links by using Perl modules \texttt{LWP::UserAgent} and \texttt{LWP::RobotUA} as in the previous study~\citep{hata20199}.
\minor{Then, we identified their HTTP status codes and considered \texttt{2xx} codes accessible links. Unlike in the previous study, we excluded error status codes (i.e., \texttt{4xx} and \texttt{5xx}) because error details are not the focus of this study.}
\minor{Finally, redirection status codes (i.e., \texttt{3xx}) occurred infrequently in the previous study---only 0.7\% of the analyzed links were redirected.
For this study, we consider such redirect links not accessible in terms of the original resource and thus excluded them.
Additional retrieval and verification of their redirection targets would be required, with a marginal effect on the results.
They could, however, be further analyzed in future work to better understand the evolution of software documentation.}

\subsection{Qualitative Analysis}
\label{sec:qa}

To understand what kinds of link targets are referenced in commit messages (\textbf{RQ2}), understand what purpose do links in commit messages serve (\textbf{RQ3}), and investigate the phenomenon of repeated links in commit messages (\textbf{RQ4}), we conduct qualitative analyses of a statistically representative and stratified sample. 

\begin{figure}[t]
  \centering
    \includegraphics[width=\textwidth]{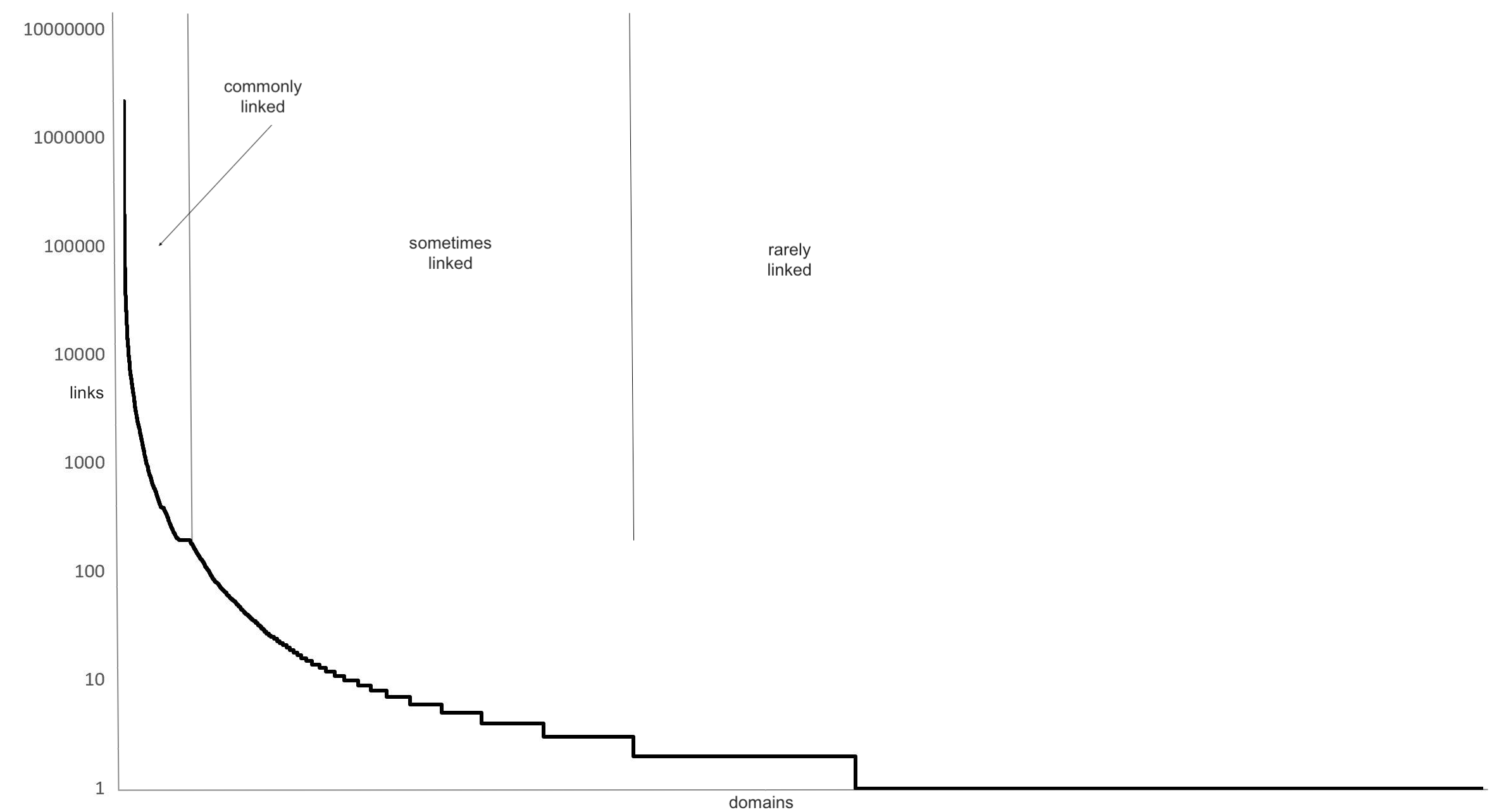}
  \caption{Distribution of links per domain.}
  \label{fig:linkperdomain}
\end{figure}

\paragraph{Link Target (RQ2)}
In RQ2, we conducted a qualitative study of a statistically representative and stratified sample of all links in our data set, to understand what kind of link targets are referenced in commit messages. Similar to the previous study of~\cite{hata20199}, we divided the data into three strata: 1) links to commonly linked domains; 2) links to domains sometimes linked; and 3) links to rarely linked domains. To decide on thresholds for distinguishing domains into three strata, we conducted a visual analysis of the distribution of links per domain in our data set. Figure~\ref{fig:linkperdomain} presents this distribution using a log scale with a cutoff frequency between `common' and `sometimes' of 194. We consider domains that account for exactly one link in our data set to be rarely linked, similar to the previous study of \cite{hata20199}.

\begin{table}[t]
\centering
\caption{Construction of the stratified sample.}
\label{tab:sampling}
\begin{tabular}{lrrrr}
\toprule
\textbf{strata} & \textbf{\# domains} & \textbf{\# links} & \textbf{\# links in sample} \\
\midrule
common & 2,270 & 17,870,470 & 384 \\
sometimes & 22,897 & 309,120 & 384 \\
rare & 21,575 & 21,575 & 377 \\
\midrule
\textbf{sum} & \textbf{46,742} & \textbf{18,201,165} & \textbf{1,145} \\
\bottomrule
\end{tabular}
\end{table}

Table~\ref{tab:sampling} shows the number of domains and the number of links in each \fix{stratum}. We randomly sampled a statistically representative number of links with a confidence level of 95\% and a confidence interval of 5 from each \fix{stratum}. In the results, we obtained 384 (from commonly linked domains), 384 (from domains sometimes linked), and 377 links (from rarely linked domains).


\cite{hata20199} introduced the coding guide for link targets in comments of source code files. In this paper, we will reuse the coding guide to classify link targets. However, we found that their coding guide did not cover all link targets in commit messages. Hence, the following three additional codes emerged from our manual analysis in the first iteration: 
\begin{itemize}
\item \fix{\textbf{repository}}: online storage location of software packages.
\item \fix{\textbf{pull request}}: request to merge changes back into the main branch.
\item \fix{\textbf{patch}}: set of changes to a repository.
\end{itemize}

The first, third, and fourth authors 
of this paper independently coded 30 links from the sample and then calculated the \fix{Fleiss's k}appa agreement~\citep{fleiss1971measuring} of this iteration between all three raters. The kappa agreement of the link target was 0.72 or ``Substantial agreement''~\citep{viera2005understanding}. Based on this encouraging result, the remaining data was then coded by the first author of this paper. \fix{If the first author was unable to identify the code of the links (i.e., 13 cases), we engage in discussions to determine the appropriate type. This collaborative approach ensured that all links were accurately and consistently coded.}

\paragraph{Link Purpose (RQ3)}

In RQ3, we use the same sample and process as RQ2. We conducted a qualitative analysis of our statistically representative sample, focusing on the purpose of the link. We also found that their coding guide~\citep{hata20199} did not cover all link purposes in commit messages. Hence, the following three additional codes emerged from our manual analysis in the first iteration:
\begin{itemize}
\item \fix{\textbf{version control sync}}: the link explicitly indicates that it is used for keeping files in sync between versions (e.g., merge branch and git-svn-id).
\item \fix{\textbf{related issue}}: the link relates to the issue report.
\item \fix{\minor{\textbf{unknown (404)}}}: exception which is the original GitHub commit link cannot be accessed. \fix{Note that creating \minor{this} code was likely not the purpose of adding the link, but since we cannot reliably access its content anymore, we code all such instances as \minor{unknown (404)}.}
\end{itemize}

After coding 30 links from the sample, the kappa agreement reached 0.69, which is also ``Substantial agreement''~\citep{viera2005understanding}. The remaining data was coded by the first author of this paper.

\paragraph{Repeated Links (RQ4)}
The corresponding research question of the previous study \citep{hata20199} is ``How do links in source code comments evolve?''
Unlike links in source code comments, it is unlikely that links in commit messages are edited in practice. Therefore, this research question, ``To what extent do commit message links get repeatedly referenced?'', investigates whether the same links appear in commit messages when the same files are modified.

We use the same sample from RQ2 and RQ3. For each link, we extract a commit history of the files modified by the commit, including the link.
Then, we count the frequency of the link in the commit history.
If the same link appears more than once, we consider the link
as repeated. We manually read the commit messages to analyze relationships among the commits. We classify them into groups based on potential reasons in a bottom-up manner.

\begin{itemize}
\item \fix{\textbf{same data source}}: includes links that appear in commits importing updates from external repositories. 
In particular, we classify a link into this group if:
\begin{itemize}
    \item the commit message is generated by a tool such as \texttt{Dependabot} and \texttt{Greenkeeper}, or
    \item the link points to a language translation platform such as \texttt{translatewiki.net} and \texttt{Weblate}.
\end{itemize}
In those cases, we consider that the links point to external data sources of the project.
\item \fix{\textbf{same purpose}}: includes the link pointing to development issues, including feature requests and bug reports. 
Links repeatedly appear in the commit history, probably because the developers needed a number of changes to resolve the issues.
We classify a link into this group if:
\begin{itemize}
    \item the link target is an issue on an issue tracking system, or
    \item commits including the link that has exactly the same message.
\end{itemize}
\item \fix{\textbf{same reference}}: includes the link that refers to either specification documents or API documents outside of the projects.
We classify a link into this group based on the link target analyzed in RQ2.
\item \fix{\textbf{other}}: includes the link that could not be categorized due to limited information.
\end{itemize}
While the codes and rules emerged from manual analysis, the sample is analyzed by the above rules.

\section{Results}
\label{sec:rs}
In this section, we present the results of each research question.

\subsection{Prevalence of Links (RQ1)}
To explore the prevalence of links referenced in commit messages, we conducted quantitative analyses of our collected data set in terms of the existence of links, diversity of domains, and popularity of domains.

\begin{figure*}[t]
    \centering
    \begin{minipage}[b]{\textwidth}
    \centering
    \subfigure[Ratio of repositories with links.\label{fig:rq1.a}]{\includegraphics[width=.49\textwidth]{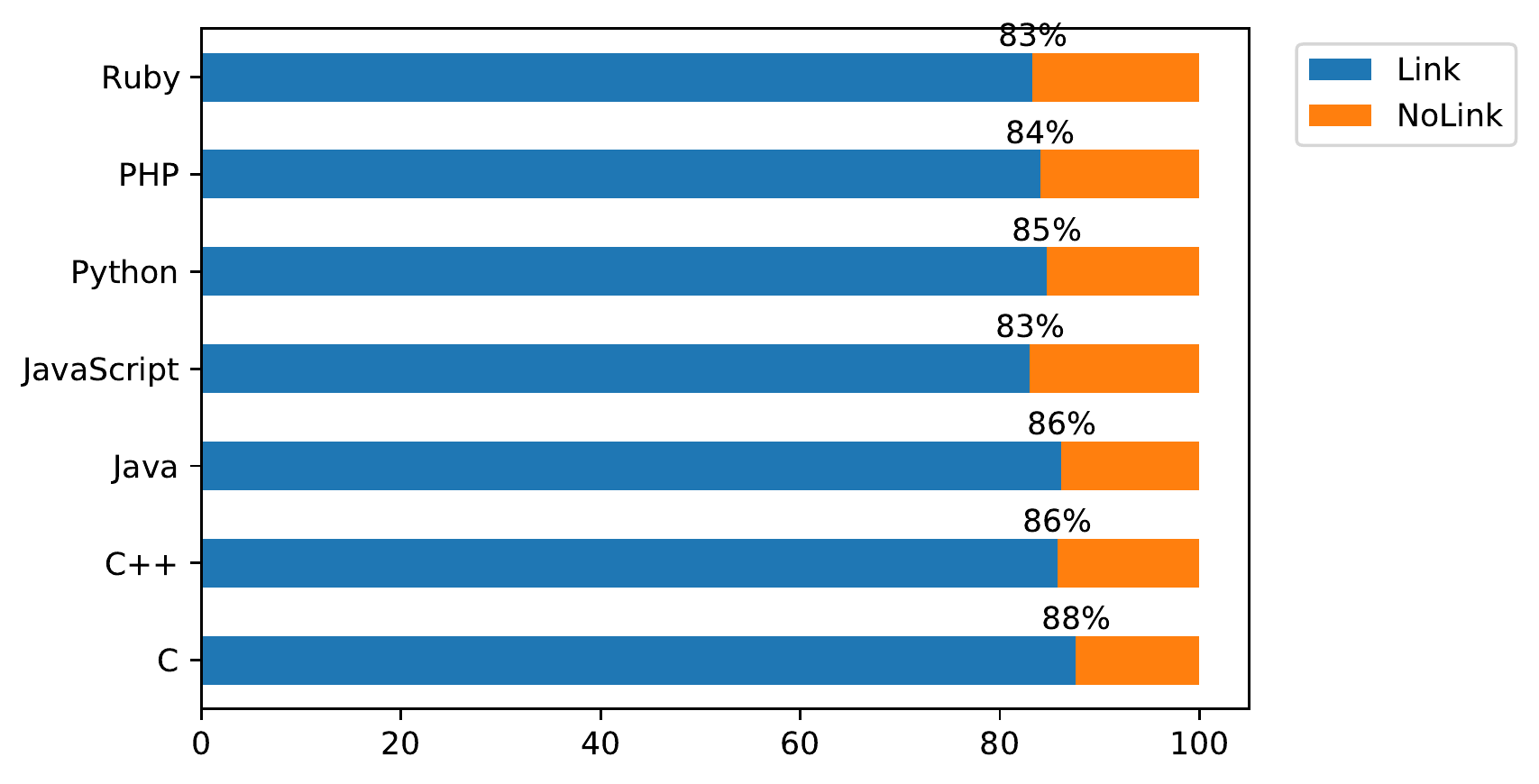}}
    \hfill
    \subfigure[Distribution of the number of different domains per repository.\label{fig:rq1.b}]{
    \includegraphics[width=.49\textwidth]{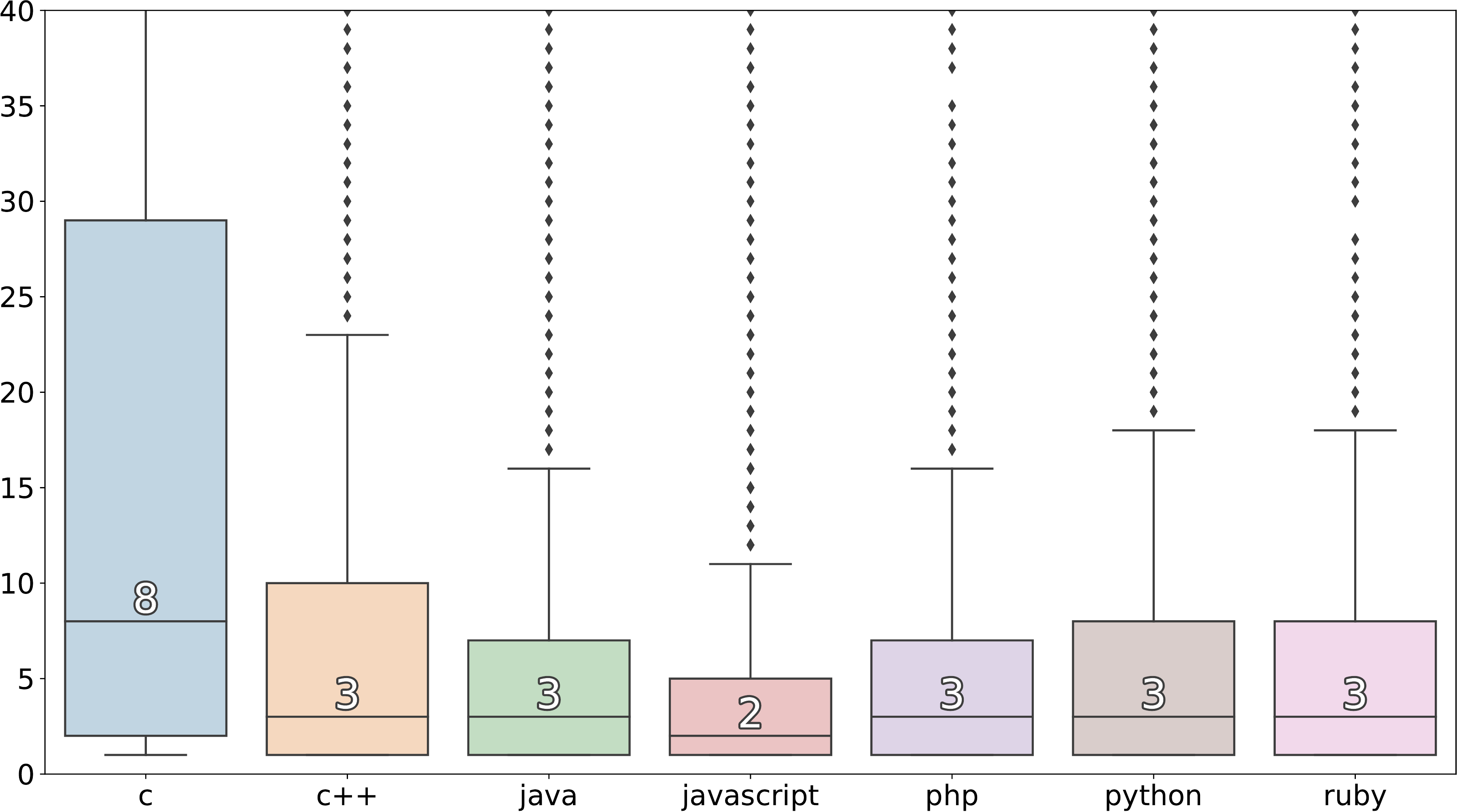}}
    \end{minipage}
    \begin{minipage}[b][]{\textwidth}
    \centering
    \subfigure[\fix{Heatmap} of repository languages shared by the top ten most referenced domains. \label{fig:rq1.c}]{
    \includegraphics[width=.79\textwidth]{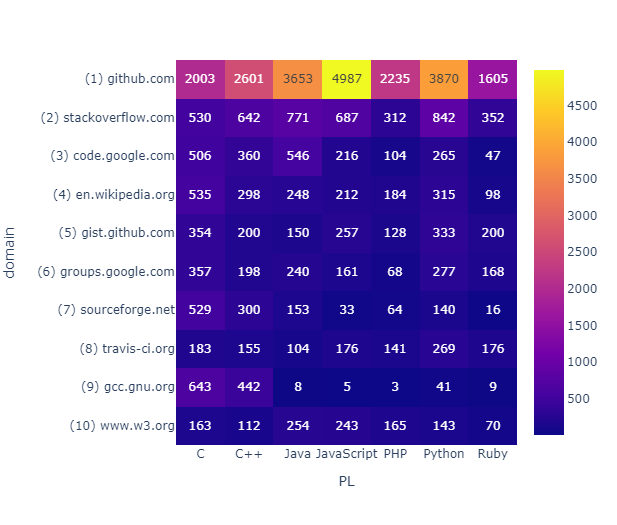}}
    
    \end{minipage}
    \caption{Analysis of links by (a) languages, (b) domain diversity, and (c) top domains.}
    \label{fig:rq1}
    \end{figure*}
    
\textbf{Link existence.}
Figure~\ref{fig:rq1.a} presents the percentages of studied repositories. Many repositories have at least one link in their commit messages, e.g., 83\% for Ruby repositories. We observe that the percentages slightly vary by language, accounting for five percentage points. Additionally, for repositories written in Java, C++, and C, more than 85\% of the repositories contain links to provide additional knowledge.

\textbf{Domain diversity.}
We obtained 46,742 distinct domains or Internet hostnames, out of the obtained 18,201,165 links. Figure~\ref{fig:rq1.b} shows the distribution of the number of different distinct domains per repository, with median values. We find that the median of different domains of repositories written in C is relatively greater than other studied languages, accounting for at least five different domains.

\textbf{Popular domains.} Figure~\ref{fig:rq1.c} depicts the \fix{heatmap} of repositories in each language shared by the top ten most referenced domains. To rank the top ten domains, we only counted once even if the domain has appeared several times, and used the number of repositories instead of the number of links.

\begin{table}[t]
\caption{\fix{Links to domain \textsf{github.com} by language. Same owner: for link pattern \lstinline{github.com/([^/]+)}, the group matches the owner of the repository the commit belongs to; same repo: in addition, the second group matches \minor{the owner name and} the repository name: \lstinline{github.com/([^/]+)/([^/]+)}.}}
\label{tab:github-internal-links}
\centering
\fix{
\begin{tabular}{lrrrrr}
\toprule
\multicolumn{1}{c}{\textbf{language}}   & \multicolumn{1}{c}{\textbf{all}} & \multicolumn{2}{c}{\textbf{same owner}} & \multicolumn{2}{c}{\textbf{same repo}} \\
\midrule
C          & 178,250      & 42,310            & (23.7\%)          & 36,225         & (20.3\%)       \\
C++        & 107,308      & 70,491            & (65.7\%)          & 54,544         & (50.8\%)       \\
Java       & 194,086      & 141,324           & (72.8\%)          & 84,502         & (43.5\%)       \\
JavaScript & 318,043      & 170,478           & (53.6\%)          & 110,255        & (34.7\%)       \\
PHP        & 116,368      & 61,632            & (53.0\%)          & 45,469         & (39.1\%)       \\
Python     & 154,938      & 84,491            & (54.5\%)          & 52,805         & (34.1\%)       \\
Ruby       & 128,686      & 28,700            & (22.3\%)          & 20,030         & (15.6\%)       \\
\midrule
\textbf{sum}        & \textbf{1,197,679}    & \textbf{599,426}           & \textbf{(50.0\%)}          & \textbf{403,830}        & \textbf{(33.7\%)}      \\
\bottomrule
\end{tabular}
}
\end{table}

\begin{table}[t]
\caption{\fix{Links to domain \textsf{github.com} pointing to the same repository (n=403,830), ten most frequent path segments following the repository name; none: link pointed to the repository root path.}}
\label{tab:github-internal-links-same-repo} 
\centering
\fix{
\begin{tabular}{lrr}
\toprule
\textbf{path segment} & \multicolumn{2}{c}{\textbf{link count}} \\
\midrule
none & 187,400 & (46.4\%) \\
issues & 107,479 & (26.6\%) \\
pull & 82,051 & (20.3\%) \\
commit & 14,667 & (3.6\%) \\
blob & 8,389 & (2.1\%) \\
compare & 1,637 & (0.4\%) \\
wiki & 983 & (0.2\%) \\
tree & 341 & (0.08\%) \\
releases & 197 & (0.05\%) \\
projects & 68 & (0.02\%) \\
\bottomrule
\end{tabular}
}
\end{table}

The \textsf{github.com} domain is the most frequently occurring domain in our studied data set, accounting for 20,954 repositories across seven languages referencing content on \textsf{github.com}. As one of the most popular social coding platforms,~\cite{dabbish2012social} found that transparency in GitHub allowed work to evolve with collaboration. Not surprisingly, the \textsf{github.com} domain is used frequently in commit messages.
\fix{Hence, we decided to look at those links in detail. Table~\ref{tab:github-internal-links} shows how many links in the particular programming languages pointed to resources hosted under the same owner as the one owning the repository the commit belongs to. The table further shows how many links pointed to resources in the same repository \minor{(matching the owner name and the repository name)}. Overall, half of the \textsf{github.com} links were internal to the repository owner, and about a third of the links were internal to the repository, i.e., pointed to resources within the same repository. Focusing on the latter, we see that almost half of those links point to the repository root path, 26.6\% point to issues, and 20.3\% point to pull requests (see Table~\ref{tab:github-internal-links-same-repo}). This indicates that almost half of the repository-internal links are used in cases where GitHub's hash notation for linking issues or pull requests (\lstinline{#issue_id} or \lstinline{#pull_request_id}) could have been used.}

Another commonly occurring domain is the \textsf{stackoverflow.com} domain, accounting for 4,136 repositories across seven languages referencing content on \textsf{stackoverflow.com}. This finding confirms the results of \cite{vasilescu2013stackoverflow}, they found Stack Overflow activities accelerate GitHub committing. Developers share external knowledge through links from Stack Overflow to GitHub to encourage committing since active developers in GitHub are also active questioners~\citep{xiong2017mining}.

The distribution of the top ten domains differs by language. But in \fix{summary}, most domains are frequently referenced in C repositories, e.g., \textsf{en.wikipe\\dia.org}, \textsf{sourceforge.net}, and \textsf{gcc.gnu.org}. The \textsf{github.com} domain is commonly referenced in \fix{4,987} JavaScript repositories. Moreover, Java repositories contain more links with the \textsf{code.google.com} domain and \textsf{www.w3.org} domain. Repositories written in C referenced many links to the domains of \textsf{en.wikipedia.org}, \textsf{gist.github.com}, \textsf{groups.google.com}, \textsf{sourceforge.net}, and \textsf{gcc.gnu.org}.

\begin{tcolorbox}
\textbf{RQ1 Summary:} 
We observe that links in commit messages are prevalent. Most of \fix{the} repositories have at least one link in their commit messages, e.g., 83\% for Ruby. The top three most frequently referenced domains per repository are \textsf{github.com}, \textsf{stackoverflow.com}, and \textsf{code.google.com}.
\end{tcolorbox}




\subsection{Link Targets (RQ2)}

\begin{table}[t]
\centering
\caption{Frequency of link target types in our sample. The bold target categories are complemented \fix{by this paper} from the \minor{previous} work~\citep{hata20199}. \minor{To prevent any confusion regarding the inclusion of ``404'' as a link target, we have decided to rename ``404''~\citep{hata20199} to ``unknown (404)''.}}
\label{tab:target}    
\resizebox{.99\textwidth}{!}{\begin{tabular}{lr@{}rr@{}rr@{}r|r@{}rr@{}rr@{}r}
\toprule
 & \multicolumn{6}{c|}{\fix{Commit messages}} & \multicolumn{6}{c}{\fix{Source code comments~\citep{hata20199}}}  \\ & \multicolumn{2}{c}{\textbf{common}} &  \multicolumn{2}{c}{\textbf{sometimes}} &  \multicolumn{2}{c|}{\textbf{rare}} &  \multicolumn{2}{c}{\fix{\textbf{common}}} &  \multicolumn{2}{c}{\fix{\textbf{sometimes}}} &  \multicolumn{2}{c}{\fix{\textbf{rare}}} \\
\midrule
\minor{unknown (404)}                                    & 161                           & (42\%)                            & 132                           & (34\%)                            & 162                           & (43\%)                            & \fix{ 27}                            & \fix{ (7\%)}                             & \fix{ 122}                           & \fix{ (32\%)}                            & \fix{ 138}                           & \fix{ (37\%)}                            \\
\textbf{patch}        & 110                           & (29\%)                            & 9                             & (2\%)                             & 2                             & (1\%)                             & \fix{ -}                             & \fix{ -}                                 & \fix{ -}                             & \fix{ -}                                 & \fix{ -}                             & \fix{ -}                                 \\
bug report                             & 61                            & (16\%)                            & 23                            & (6\%)                             & 2                             & (1\%)                             & \fix{ 9}                             & \fix{ (2\%)}                             & \fix{ 10}                            & \fix{ (3\%)}                             & \fix{ 3}                             & \fix{ (1\%)}                             \\
other                                  & 20                            & (5\%)                             & 49                            & (13\%)                            & 51                            & (14\%)                            & \fix{ 5}                             & \fix{ (1\%)}                             & \fix{ 23}                            & \fix{ (6\%)}                             & \fix{ 45}                            & \fix{ (12\%)}                            \\
\textbf{repository}   & 14                            & (4\%)                             & 2                             & (1\%)                             & 1                             & (0\%)                             & \fix{ -}                             & \fix{ -}                                 & \fix{ -}                             & \fix{ -}                                 & \fix{ -}                             & \fix{ -}                                 \\
software homepage                      & 6                             & (2\%)                             & 26                            & (7\%)                             & 23                            & (6\%)                             & \fix{ 55}                            & \fix{ (14\%)}                            & \fix{ 65}                            & \fix{ (17\%)}                            & \fix{ 28}                            & \fix{ (7\%)}                             \\
organization homepage                  & 4                             & (1\%)                             & 13                            & (3\%)                             & 15                            & (4\%)                             & \fix{ 16}                            & \fix{ (4\%)}                             & \fix{ 41}                            & \fix{ (11\%)}                            & \fix{ 24}                            & \fix{ (6\%)}                             \\
tutorial or article                    & 3                             & (1\%)                             & 34                            & (9\%)                             & 42                            & (11\%)                            & \fix{ 16}                            & \fix{ (4\%)}                             & \fix{ 21}                            & \fix{ (5\%)}                             & \fix{ 31}                            & \fix{ (8\%)}                             \\
\textbf{pull request} & 3                             & (1\%)                             & 1                             & (0\%)                             & 0                             & (0\%)                             & \fix{ -}                             & \fix{ -}                                 & \fix{ -}                             & \fix{ -}                                 & \fix{ -}                             & \fix{ -}                                 \\
forum thread                           & 2                             & (1\%)                             & 28                            & (7\%)                             & 4                             & (1\%)                             & \fix{ 0}                             & \fix{ (0\%)}                             & \fix{ 5}                             & \fix{ (1\%)}                             & \fix{ 6}                             & \fix{ (2\%)}                             \\
API documentation                      & 0                             & (0\%)                             & 22                            & (6\%)                             & 12                            & (3\%)                             & \fix{ 14}                            & \fix{ (4\%)}                             & \fix{ 20}                            & \fix{ (5\%)}                             & \fix{ 10}                            & \fix{ (3\%)}                             \\
blog post                              & 0                             & (0\%)                             & 14                            & (4\%)                             & 23                            & (6\%)                             & \fix{ 1}                             & \fix{ (0\%)}                             & \fix{ 10}                            & \fix{ (3\%)}                             & \fix{ 22}                            & \fix{ (6\%)}                             \\
specification                          & 0                             & (0\%)                             & 12                            & (3\%)                             & 4                             & (1\%)                             & \fix{ 21}                            & \fix{ (5\%)}                             & \fix{ 33}                            & \fix{ (9\%)}                             & \fix{ 32}                            & \fix{ (8\%)}                             \\
application                            & 0                             & (0\%)                             & 7                             & (2\%)                             & 15                            & (4\%)                             & \fix{ 0}                             & \fix{ (0\%)}                             & \fix{ 11}                            & \fix{ (3\%)}                             & \fix{ 13}                            & \fix{ (3\%)}                             \\
code                                   & 0                             & (0\%)                             & 5                             & (1\%)                             & 7                             & (2\%)                             & \fix{ 6}                             & \fix{ (2\%)}                             & \fix{ 2}                             & \fix{ (1\%)}                             & \fix{ 5}                             & \fix{ (1\%)}                             \\
research paper                         & 0                             & (0\%)                             & 4                             & (1\%)                             & 4                             & (1\%)                             & \fix{ 0}                             & \fix{ (0\%)}                             & \fix{ 9}                             & \fix{ (2\%)}                             & \fix{ 13}                            & \fix{ (3\%)}                             \\
personal homepage                      & 0                             & (0\%)                             & 2                             & (1\%)                             & 6                             & (2\%)                             & \fix{ 4}                             & \fix{ (1\%)}                             & \fix{ 8}                             & \fix{ (2\%)}                             & \fix{ 4}                             & \fix{ (1\%)}                             \\
Q\&A thread                            & 0                             & (0\%)                             & 1                             & (0\%)                             & 1                             & (0\%)                             & \fix{ 0}                             & \fix{ (0\%)}                             & \fix{ 0}                             & \fix{ (0\%)}                             & \fix{ 1}                             & \fix{ (0\%)}                             \\
license                                & 0                             & (0\%)                             & 0                             & (0\%)                             & 2                             & (1\%)                             & \fix{ 208}                           & \fix{ (54\%)}                            & \fix{ 4}                             & \fix{ (1\%)}                             & \fix{ 1}                             & \fix{ (0\%)}                             \\
book content                           & 0                             & (0\%)                             & 0                             & (0\%)                             & 1                             & (0\%)                             & \fix{ 0}                             & \fix{ (0\%)}                             & \fix{ 0}                             & \fix{ (0\%)}                             & \fix{ 2}                             & \fix{ (1\%)}                             \\
GitHub profile                         & 0                             & (0\%)                             & 0                             & (0\%)                             & 0                             & (0\%)                             & \fix{ 1}                             & \fix{ (0\%)}                             & \fix{ 0}                             & \fix{ (0\%)}                             & \fix{ 0}                             & \fix{ (0\%)}                             \\
Stack Overflow                         & 0                             & (0\%)                             & 0                             & (0\%)                             & 0                             & (0\%)                             & \fix{ 1}                             & \fix{ (0\%)}                             & \fix{ 0}                             & \fix{ (0\%)}                             & \fix{ 0}                             & \fix{ (0\%)}                             \\
\midrule
\textbf{sum}          & \textbf{384} & \textbf{(100\%)} & \textbf{384} & \textbf{(100\%)} & \textbf{377} & \textbf{(100\%)} & \fix{ \textbf{384}} & \fix{ \textbf{(100\%)}} & \fix{ \textbf{384}} & \fix{ \textbf{(100\%)}} & \fix{ \textbf{378}} & \fix{ \textbf{(100\%)}} \\ 
\bottomrule
\end{tabular}}
\end{table}


Table \ref{tab:target} shows the result of our qualitative analysis. For all types of domains, \minor{in many of the cases we could not determine
the link target type because of a missing link}, accounting for 42\%, 34\%, and 43\%, respectively. For commonly-linked domains, \texttt{patch} is the second most frequent type of link target, accounting for 29\%. For domains that are sometimes and rarely linked, \texttt{tutorial or article} is the third most common type of link target. Moreover, this table reveals that the remaining link targets are distributed similarly (i.e., not more than 7\%). The prevalence of the code \texttt{other} in the results for links to
all linked domains is an indicator of the diversity of links
present in commit messages.




\begin{tcolorbox}
\textbf{RQ2 Summary:} 
\minor{Inaccessible} links are the most prevalent target type in commit messages, whereas various other types,
such as \texttt{patch}, \texttt{bug report}, and \texttt{tutorial or article}, are also common.
\end{tcolorbox}

\subsection{Link Purpose (RQ3)}

\begin{table}[t]
\centering
\caption{Frequency of link purposes in our sample. The bold target categories are complemented \fix{by this paper} from the \minor{previous} work~\citep{hata20199}.}
\label{tab:purpose}      
\resizebox{.99\textwidth}{!}{\begin{tabular}{lr@{}rr@{}rr@{}r|r@{}rr@{}rr@{}r}
\toprule
 & \multicolumn{6}{c|}{\fix{Commit messages}} & \multicolumn{6}{c}{\fix{Source code comments~\citep{hata20199}}}  \\ & \multicolumn{2}{c}{\textbf{common}} &  \multicolumn{2}{c}{\textbf{sometimes}} &  \multicolumn{2}{c|}{\textbf{rare}} &  \multicolumn{2}{c}{\fix{\textbf{common}}} &  \multicolumn{2}{c}{\fix{\textbf{sometimes}}} &  \multicolumn{2}{c}{\fix{\textbf{rare}}} \\
\midrule
\textbf{version control sync} & 169                           & (44\%)                            & 18                            & (5\%)                             & 1                             & (0\%)                             & \fix{-}                             & \fix{-}                                 & \multicolumn{1}{r|}{\fix{-}}                             & \fix{-}                                 & \fix{-}                             & \fix{-}                                 \\
metadata                                       & 85                            & (22\%)                            & 11                            & (3\%)                             & 12                            & (3\%)                             & \fix{288}                           & \fix{(75\%)}                            & \multicolumn{1}{r|}{\fix{131}}                           & \fix{(34\%)}                            & \fix{43}                            & \fix{(11\%)}                            \\
\textbf{related issue}        & 70                            & (18\%)                            & 54                            & (14\%)                            & 12                            & (3\%)                             & \fix{-}                             & \fix{-}                                 & \multicolumn{1}{r|}{\fix{-}}                             & \fix{-}                                 & \fix{-}                             & \fix{-}                                 \\
source code context                            & 28                            & (7\%)                             & 225                           & (59\%)                            & 267                           & (71\%)                            & \fix{18}                            & \fix{(5\%)}                             & \multicolumn{1}{r|}{\fix{60}}                            & \fix{(16\%)}                            & \fix{80}                            & \fix{(21\%)}                            \\
source/attribution                             & 16                            & (4\%)                             & 44                            & (11\%)                            & 62                            & (16\%)                            & \fix{27}                            & \fix{(7\%)}                             & \multicolumn{1}{r|}{\fix{62}}                            & \fix{(16\%)}                            & \fix{75}                            & \fix{(20\%)}                            \\
\minor{\textbf{unknown (404)}}                  & 11                            & (3\%)                             & 2                             & (1\%)                             & 1                             & (0\%)                             & \fix{-}                             & \fix{-}                                 & \multicolumn{1}{r|}{\fix{-}}                             & \fix{-}                                 & \fix{-}                             & \fix{-}                                 \\
see-also                                       & 3                             & (1\%)                             & 14                            & (4\%)                             & 12                            & (3\%)                             & \fix{28}                            & \fix{(7\%)}                             & \multicolumn{1}{r|}{\fix{59}}                            & \fix{(15\%)}                            & \fix{51}                            & \fix{(13\%)}                            \\
commented-out source code                      & 1                             & (0\%)                             & 11                            & (3\%)                             & 9                             & (2\%)                             & \fix{1}                             & \fix{(0\%)}                             & \multicolumn{1}{r|}{\fix{17}}                            & \fix{(4\%)}                             & \fix{70}                            & \fix{(19\%)}                            \\
link-only                                      & 1                             & (0\%)                             & 2                             & (1\%)                             & 1                             & (0\%)                             & \fix{6}                             & \fix{(2\%)}                             & \multicolumn{1}{r|}{\fix{24}}                            & \fix{(6\%)}                             & \fix{40}                            & \fix{(11\%)}                            \\
self-admitted technical debt                   & 0                             & (0\%)                             & 2                             & (1\%)                             & 0                             & (0\%)                             & \fix{11}                            & \fix{(3\%)}                             & \multicolumn{1}{r|}{\fix{16}}                            & \fix{(4\%)}                             & \fix{13}                            & \fix{(3\%)}                             \\
@see                                           & 0                             & (0\%)                             & 1                             & (0\%)                             & 0                             & (0\%)                             & \fix{5}                             & \fix{(1\%)}                             & \multicolumn{1}{r|}{\fix{15}}                            & \fix{(4\%)}                             & \fix{6}                             & \fix{(2\%)}                             \\ \midrule
\textbf{sum}                  & \textbf{384} & \textbf{(100\%)} & \textbf{384} & \textbf{(100\%)} & \textbf{377} & \textbf{(100\%)} & \fix{\textbf{384}} & \fix{\textbf{(100\%)}} & \multicolumn{1}{r|}{\fix{\textbf{384}}} & \fix{\textbf{(100\%)}} & \fix{\textbf{378}} & \fix{\textbf{(100\%)}} \\ 
\bottomrule
\end{tabular}}
\end{table}

Table~\ref{tab:purpose} shows the result of our qualitative analysis. For commonly-linked domains, \texttt{version control sync} is the most frequent purpose, accounting for 44\% of links, followed by \texttt{metadata} (22\%). For domains that are sometimes and rarely linked, \texttt{source code context} is the most common purpose (59--71\%), followed by \texttt{source/attribution} (11--16\%) and \texttt{related issue} (3--14\%), respectively. This indicates that most links are used for keeping files in sync or adding additional information to fill the context in commit messages. We also observe a few false-positive cases (0--3\%), which indicate the original GitHub commit link cannot be accessed.


\textbf{Patterns in the relationship between link targets and purposes.} Based on the qualitative analysis conducted to answer RQ2 and RQ3 about
the targets and purposes of links in source code comments,
we can now investigate the relationships between the
different types of link targets and the different purposes which emerged from our qualitative analysis. To do so, we applied
association rule learning using the apriori algorithm~\citep{agrawal1994fast} as implemented in the R package arules\footnote{\url{https://cran.r-project.org/web/packages/arules/index.html}} to our data, treating
each link as a transaction containing two items: its target type
and its purpose. We used four as the support threshold and 0.7
as the confidence threshold, i.e., all the rules that we extracted are
supported by at least four data points, and we have at least a
70\% confidence that the left-hand side of the rule implies the
right-hand side.

\begin{table}[t]
\centering
\caption{Associations between link target type and link purpose.}
\label{tab:rule}      
\begin{tabular}{llclrr}
\toprule
\textbf{strata}&  \multicolumn{3}{c}{\textbf{ association rule}} &  \textbf{conf.} &  \textbf{supp.} \\
\midrule
common & \minor{unknown (404)}  & $\Rightarrow$ & version control sync  & 0.85 & 132 \\
common & version control sync & $\Rightarrow$ & \minor{unknown (404)} & 0.78 & 132 \\ 
common & metadata & $\Rightarrow$ & patch & 0.85 & 72 \\ 
common & bug report & $\Rightarrow$ & related issue & 0.97 & 56 \\ 
common & related issue & $\Rightarrow$ & bug report & 0.80 & 56 \\ 
\midrule
sometimes & software homepage & $\Rightarrow$ & source code context & 0.81 & 21 \\ 
sometimes & bug report & $\Rightarrow$ & related issue & 0.87 & 20 \\ 
sometimes & version control sync & $\Rightarrow$ & \minor{unknown (404)} & 0.94 & 17 \\ 
sometimes & organization homepage & $\Rightarrow$ & source code context & 0.85 & 11 \\ 
sometimes & application & $\Rightarrow$ & source code context & 0.86 & 6 \\ 
\midrule
rare & \minor{unknown (404)} & $\Rightarrow$ & source code context & 0.77 & 125 \\ 
rare & other & $\Rightarrow$ & source code context & 0.80 & 41 \\ 
rare & software homepage & $\Rightarrow$ & source code context & 0.86 & 19 \\ 
rare & application & $\Rightarrow$ & source code context & 1.00 & 15 \\ 
rare & API documentation & $\Rightarrow$ & source code context & 0.75 & 9 \\ 
rare & commented-out source code & $\Rightarrow$ & \minor{unknown (404)} & 0.89 & 8 \\ 
rare & forum thread & $\Rightarrow$ & source code context & 1.00 & 4 \\ 
\bottomrule
\end{tabular}
\end{table}

Table~\ref{tab:rule} shows the association rules extracted from our data
with these settings, separately for each \fix{stratum} in our sample. \minor{Especially in commonly-linked domains, we observe a tight connection between keeping files in sync and links no longer being available, i.e., \texttt{unknown (404)}}. 85\% of the \minor{inaccessible} links are \fix{related to} the purpose of keeping files in sync, and the 78\% of links that were added in commit messages for the purpose of keeping files in sync are found to be inaccessible. Other than \minor{inaccessible} links being tightly connected with the purpose of keeping files in sync, we identified more relationships (e.g., between bug report links and the purpose of providing related issues, and links to software, organization, application, API documentation, and forum thread are associated with source code context).

\fix{After examining commit messages containing 171 Subversion-related links (links containing `svn'), we found that five cases were inaccessible on GitHub, 164 messages contained the keyword `git-svn-id', and two contained the keyword `svnmerge'. The `git-svn-id' is a keyword that appears when migrating from Subversion to Git, so the fact that this link is currently inaccessible does not mean that any important information is missing. The keyword `svnmerge' is related to merging in Subversion, and not being able to access this link can be problematic for understanding merge details. However, the number of such commit messages is small.}

\begin{tcolorbox}
\textbf{RQ3 Summary:}
For domains that are sometimes and rarely linked, \texttt{source code context} is the most prevalent purpose in commit messages. The purpose \texttt{version control sync} is particularly common for commonly linked domains.
\end{tcolorbox}

\subsection{Repeated Links (RQ4)}

\begin{table}[t]
\centering
\caption{Categories of links that repeatedly appear in the commit history.}
\label{tab:repeated}      
\begin{tabular}{lr@{}rr@{}rr@{}rr}
\toprule
Category 
&  \multicolumn{2}{c}{\textbf{common}} &  \multicolumn{2}{c}{\textbf{sometimes}} &  \multicolumn{2}{c}{\textbf{rare}} &  
frequency \\
\midrule
same data source    & 12 & (3.1\%) & 6 & (1.6\%) & 0 & (0\%) & 2--260 \\  
same purpose        &  9 & (2.3\%) & 8 & (2.1\%)    & 0 & (0\%) & 2--8 \\
same reference      &  0 & (0.0\%) & 4 & (1.0\%) & 0 & (0\%) & 2--5 \\
other               &  1 & (0.3\%) & 6 & (1.6\%)    & 0 & (0\%) & 2--28 \\
\midrule
\textbf{sum} & \textbf{22} & \textbf{(5.7\%)} & \textbf{24} & \textbf{(6.3\%)} & \textbf{0} & \textbf{(0\%)} & \textbf{-}\\
\bottomrule
\end{tabular}
\end{table}


Table~\ref{tab:repeated} shows the number of categories we have identified in repeated links.
We observed that only 5.7\% and 6.3\% of sample from commonly linked domains and domains sometimes linked appear more than once in their commit histories and no link repeatedly appears in rarely linked domains.
The result shows that, in most cases, developers use external documents only once to complete a task.
For commonly linked domains, \texttt{same data source} is the most frequent reason, accounting for 12 links, followed by \texttt{same purpose} (9). For domains that are sometimes linked, \texttt{same purpose} is the most common reason, accounting for 8 links, followed by \texttt{same data source} (6). 
Besides that, the repeated links due to \texttt{same data source} could vary from 2--260 times. 
The most frequent link is pointed to \textsf{translatewiki.net}. 
The result indicates that developers continuously use the platform to update their documentation.

This analysis examined links in commit messages when the same files were edited, whereas the previous study \citep{hata20199} examined how links in code comments were edited. Therefore, the results are not comparable.



\begin{tcolorbox}
\textbf{RQ4 Summary:}
Repeatedly referenced links in commit messages occur infrequently. Four percent of our sample is repeatedly referenced in the history of commit messages. The \texttt{same data source} is the most common reason for developers repeatedly referencing links in commit messages. The frequency of the links being repeatedly referenced varies from 2--260 times for \texttt{same data source}.
\end{tcolorbox}

\begin{table}[t]
\centering
\caption{Evolution of the link status of our sample in Wayback Machine and our coding in RQ2.}
\label{tab:evo}      
\begin{tabular}{llllr@{}r}
\toprule
\textbf{20200316} & \textbf{20210316} & \textbf{20220316} & \textbf{link target in RQ2} &  \textbf{\# of links} \\
\midrule
200 & 200 & 200	& Not \minor{unknown (404)} & 389 & (34\%) \\ 
200 & 200 & 200 & \minor{unknown (404)} & 128 & (11\%) \\
200	& 200 & Not Found & \minor{unknown (404)} & 1 & (0\%) \\
200 & Not Found	& 200 & \minor{unknown (404)}	& 3 & (0\%) \\
Not Found & 200 & 200 & \minor{unknown (404)} & 1 & (0\%) \\
200	& 200 & Not Found & Not \minor{unknown (404)} & 4 & (0\%) \\
200	& Not Found	& 200 & Not \minor{unknown (404)} & 3 & (0\%) \\
Not Found & 200 & 200 & Not \minor{unknown (404)} & 4 & (0\%) \\ 
Not Found & Not Found &Not Found & \minor{unknown (404)} & 322 & (28\%) \\ 
Not Found & Not Found & Not Found & Not \minor{unknown (404)} & 290 & (25\%) \\
\midrule
\textbf{sum} & & & &\textbf{1145} & (100\%)\\
\bottomrule
\end{tabular}
\begin{tablenotes}
      \footnotesize
      \item Not \minor{unknown (404)} represents the link targets that are not identified as \minor{unknown (404)} in our coding. Not Found represents the links that are unavailable or not archived in Wayback Machine.
\end{tablenotes}
\end{table}

\subsection{Link Target Evolution (RQ5)}
To investigate the evolution of link targets, we conducted a quantitative analysis of our sample. Table~\ref{tab:evo} shows the results of this analysis. Approximately a third of the links in our sample exist permanently, representing 34\% of our sample. Among these permanent links, \texttt{tutorial or article} (67), \texttt{other} (54), and \texttt{software homepage} (47) are the most frequent link targets. In detail, most link targets are available that are considered official (e.g., 82\% of \texttt{repository} links, 85\% of \texttt{software homepage} links, 86\% of \texttt{application} links, 88\% of \texttt{personal homepage} links, and 94\% of \texttt{organization homepage} links) or documentation (e.g., 85\% of \texttt{tutorial or article} links, 86\% of \texttt{blog post} links, 88\% of \texttt{API documentation} links, 100\% of \texttt{book content} links, 100\% of \texttt{license} links, and 100\% of \texttt{Q\&A thread} links) \fix{w}eb pages. In contrast, we find the minority of link targets are available that are considered as temporarily available software artifacts, e.g., only 9\% of \texttt{patch} links, 25\% of \texttt{pull request} links, 38\% of \texttt{research paper} links, 41\% of \texttt{bug report} links, and 41\% of \texttt{forum thread} links.

We further investigate a total of 133 links (the third row to the sixth row of Table~\ref{tab:evo}) that exist at least in one closest snapshot but we identify as \minor{\texttt{unknown (404)}} in RQ2. We code link targets of these links in Wayback snapshots following our coding guide in RQ2. We find that 24 links are still coded as \minor{\texttt{unknown (404)}}, accounting for 18\% of these links. In addition to that, \texttt{tutorial or article} and \texttt{software homepage} are frequently occurring link targets, accounting for 20 links (15\%) and 19 links (14\%), respectively. 
We also code 11 links (the seventh row to the ninth row of Table~\ref{tab:evo}) that have at least existed in one closest snapshot and we identified them as \minor{\texttt{Not unknown (404)}} in RQ2. We observe that the link targets are not changed.

28\% of the links in our sample are unavailable or not archived in snapshots and our coding. Besides that, a quarter of links in our sample are unavailable or are not archived in snapshots but are not identified as \minor{\texttt{unknown (404)}} in our coding. To investigate whether these 290 links are unavailable or are not archived in snapshots, we attempted to request these 290 links with a maximum of ten retries. We find that only 24 of 290 links are unavailable (i.e., 15 links are identified as \texttt{other} in RQ2). Thus, we find 14\% of links ($\frac{24+133}{1145} \approx14\%$) that evolve and become unavailable over time.

These results indicate that links in commit messages are fragile, unstable, and easily inaccessible; these links could cause knowledge loss from this communication channel between commit authors and developers.

To investigate the evolution of link targets in more detail, we conducted one quantitative case study with the subset of links pointing to Stack Overflow.
Of the 18,201,165 links, there are 9,696 links pointing to 7,315 distinct link targets on \textsf{stackoverflow.com}. Among those Stack Overflow links, there are varieties of expressions, e.g., an abbreviated path to an answer (\texttt{/a/(answer id)}), an abbreviated path to a question (\texttt{/q/(question id)}), and a full path to a question (\texttt{/questions/(question id)/(title)}). Older links start with `\texttt{http://}' and newer links start with `\texttt{https://}'.


We manually removed 122 commit-link pairs where the link was not referring to a specific Stack Overflow thread (e.g., Stack Overflow homepage), which left us with links to 6,973 unique post IDs. For each Stack Overflow link, we identified the timestamp of when the link was referenced in a commit message.
For duplicate links, we consider only the oldest timestamp, leaving us with 6,973 distinct links and the oldest commit that referred to them. We created a statistically representative random sample of 364 links (confidence interval: 5\%; confidence level: 95\%) of this data for our analysis.\footnote{\url{https://www.surveysystem.com/sscalc.htm}} The calculation of
statistically significant sample sizes based on population size,
confidence interval, and confidence level is well established by~\cite{krejcie1970determining}. We excluded three commits that were no longer available on GitHub at the time of analysis.

\begin{figure}[t]
\centering
\includegraphics[width=\linewidth]{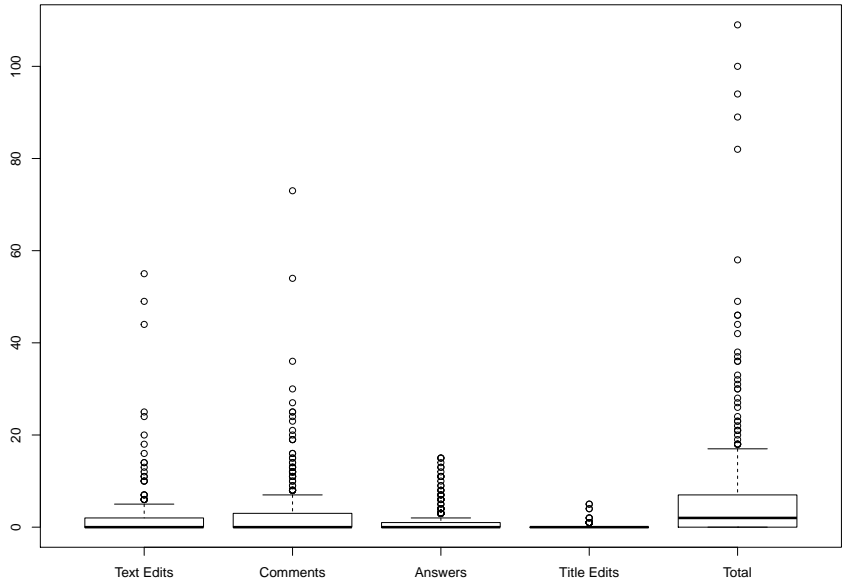}
\caption{Number of changes to the linked Stack Overflow threads after they were first referenced.}
\label{fig:so-case}
\end{figure}

We then used the SOTorrent data set~\citep{baltes2018sotorrent} to investigate the extent to which Stack Overflow content had changed since the link to the question or answer had been referenced in a commit message. We queried SOTorrent to determine the following metrics for each link:
\begin{itemize}
    \item the number of text edits on any post (i.e., question or answer) in the same thread,
    \item the number of new comments on any post (i.e., question or answer) in the same thread,
    \item the number of new answers in the same thread, and
    \item the number of edits to the thread title.
\end{itemize}

Figure~\ref{fig:so-case} shows the results of this analysis. More than half of all Stack Overflow threads experienced at least one change (median: 2, third quartile: 7) after they were referenced in a commit message, and more than a quarter of these links attracted at least one new comment in the meantime (median: 0, third quartile: 3). Although the number of new answers to a thread was zero in the median case, a quarter of the Stack Overflow threads attracted at least one new answer after the link was referenced in a commit message (median: 0, third quartile: 1). In total, only 142 (39\%) of the 361 Stack Overflow threads in our sample did not undergo any change after being included in a commit message.

The most extreme example in our sample---a Stack Overflow thread titled ``C++ Singleton design pattern''\footnote{\url{https://stackoverflow.com/q/1008019}}---had 11 new answers added (out of a total of 24) and attracted 73 new comments and 25 edits after being first referenced in a commit message in early July 2014.

\begin{tcolorbox}
\textbf{RQ5 Summary:}
In summary, 14\% of links are prone to evolve and become unavailable over time.
Moreover, in a case study of Stack Overflow, we find that more than half of Stack Overflow threads linked in commit messages attracted at least one change (text edit, new comment, new answer, or title edit) after they were referenced.
\end{tcolorbox}

\subsection{Link Decay (RQ6)}
We identified 4,659,236 \minor{inaccessible} links from the \fix{w}eb content of the 6,667,207 unique links (70\%).
Table~\ref{tab:deadlinks} shows the domains with more than 30,000 \minor{inaccessible} links. The \textsf{svn.apache.org} is the domain that occurs most frequently in these \minor{inaccessible} links, accounting for 585,149 of the 607,873 links in commit messages (96\%). 

Upon closer inspection, most domains with a large number of \minor{inaccessible} links are related to SVN (SubVersioN) repositories; for instance, \textsf{svn.apache.org} is the domain for Apache Software Foundation Subversion Server. This result confirms our finding in RQ3 and RQ5, which \minor{inaccessible} links and the purpose of keeping files in sync are tightly connected in our sample (RQ3), and patch is temporarily available software artifacts (RQ5). The \textsf{gerrit.instructure.com} domain is used for Gerrit Code Review, which requires sign-in to be accessed. Besides that, the links of \textsf{github.com} domain are unable to be accessed since they are no longer
available or pointing to private repositories. Moreover, only 8\% of the links belonging to \textsf{github.com} domain are \minor{inaccessible} in our data set.

Most of the \minor{inaccessible} links in commit messages belong to domains for SVN repositories (e.g., \textsf{svn.apache.org}) instead of \textsf{github.com} domain in source code comments. These SVN repository domains have a high chance to be \minor{inaccessible}, i.e., at least 96\%. \fix{We also observed these SVN-related links are \minor{inaccessible} in RQ3, usually these links are introduced in commit messages with the keyword `git-svn-id' that appears in the
transition from Subversion to Git, so the fact that this link is currently inaccessible does not mean
that any important information is missing.}


\begin{table}[t]
\centering
\caption{Domains with large number of \minor{inaccessible} links}
\label{tab:deadlinks}
\begin{tabular}{lrrr}
\toprule
\textbf{domain} & \textbf{\# \minor{inaccessible} links} & \textbf{total links} & \textbf{\% \minor{inaccessible} links} \\
\midrule
svn.apache.org & 585,149 & 607,873 & (96\%) \\
llvm.org & 408,888 & 410,396 & (99\%) \\
svn.code.sf.net & 370,339 & 370,351 & (99\%) \\
www.virtualbox.org & 159,144 & 159,178 & (99\%) \\
svn.osgeo.org & 89,719 & 89,721 & (99\%) \\
anonsvn.ncbi.nlm.nih.gov & 76,725 & 76,725 & (100\%) \\
root.cern.ch & 75,289 & 75,568 & (99\%) \\
svn.wxwidgets.org & 68,954 & 68,954 & (100\%) \\
svn.aros.org & 52,884 & 52,884 & (100\%) \\
dev.geogebra.org & 46,975 & 46,988 & (99\%) \\
svn.blender.org & 46,030 & 46,067 & (99\%) \\
develop.svn.wordpress.org & 41,321 & 41,325 & (99\%) \\
core.svn.wordpress.org & 41,213 & 41,213 & (100\%) \\
svn.codehaus.org & 41,202 & 41,202 & (100\%) \\
svn.parrot.org & 39,973 & 39,973 & (100\%) \\
svn.erp5.org & 38,292 & 38,322 & (99\%) \\
source.sakaiproject.org & 36,614 & 36,684 & (99\%) \\
v8.googlecode.com & 36,013 & 36,013 & (100\%) \\
gerrit.instructure.com & 34,884 & 36,708 & (95\%) \\
svn.opendtect.org & 33,977 & 33,977 & (100\%) \\
svn.jboss.org & 33,799 & 33,799 & (100\%) \\
github.com & 33,398 & 405,784 & (8\%) \\
\bottomrule
\end{tabular}
\end{table}

\begin{tcolorbox}
\textbf{RQ6 Summary:}
We observe that 70\% of links are not available, considering all unique links. The most frequently occurring domains accounting for \minor{inaccessible} links are related to SVN repositories.
\end{tcolorbox}

\section{Discussion}
\label{sec:ds}
In this section, we first compare the different roles that links play between source code comments and commit messages. Second, we make recommendations to developers and researchers based on our results.

\subsection{\fix{Comparisons of the roles} of links between source code comments and commit messages}
\fix{Based on our results, we make the following comparison in each RQ.}

\fix{\textbf{Prevalence of Links (RQ1).}
On the one hand, we find that the average ratio of repositories with links decreases from \fix{89}\%~\citep{hata20199} to 85\% in Figure~\ref{fig:rq1.a}. Especially, PHP has the highest ratio of repositories with links in source code comments (92\%), has decreased to the third lowest ratio of repositories with links in commit messages (84\%) in Figure~\ref{fig:rq1.a}, accounting for eight percentage points.}

\fix{On the other hand, we obtained around two times as many links as in source code comments (9.6 million links). Furthermore, domain diversity in commit messages (46,742 distinct domains of 18 million links) is lower than links in source code comments (57,039 distinct domains of 9.6 million links). The median values of different domains per repository have decreased by a margin of 2--10 domains in Figure~\ref{fig:rq1.b}. The \textsf{code.google.com} domain is more frequently referenced in commit messages (from 7th rank to 3rd rank in Figure~\ref{fig:rq1.c}). The ranks of licensing-related domains decreased in commit messages, e.g., \textsf{www.apache.org} (4th rank in source code comments) and \textsf{www.gnu.org} (6th rank in source code comments). However, the ranks of committing related domains increased, e.g., \textsf{travis-ci.org}.}

\fix{These comparison results indicate that developers tend to reference more links in commit messages to provide commit-related information to reviewers or other developers, however, the diversity of domains is lower than for links in source code comments.}

\fix{\textbf{Link Targets (RQ2).}
Table~\ref{tab:target} shows the results of link types in commit messages and source code comments. The decay of links (inaccessible) in commit messages is worse than in source code comments. In detail, the \minor{inaccessible} links have increased from 7--37\% to 34--43\%. Even for the domains that are commonly linked, link decay has become a common issue in commit messages (from 7\% in source code comments to 42\% in commit messages).}

\fix{Due to the different nature of code comments and commit messages, the license information has changed from the majority of link targets in source code comments (0--54\%) to the minority of link targets in commit messages (0--1\%). This finding confirms the result of RQ1, that the license information domains are not in the top ten popular domains in commit messages. On the one hand, there are two link targets (\texttt{GitHub profile} and \texttt{Stack Overflow}), which are not found in commit messages. On the other hand, we also find three new link targets, such as \texttt{repository}, \texttt{pull request}, and \texttt{patch}.}

\fix{These comparison results indicate that developers tend to reference links that are temporarily available in commit messages. These links are short-lived in the commit process even though they are linked to common domains. In commit messages, developers are more likely to reference commit process-related links, e.g., bug reports, pull requests, and patches.}


\fix{\textbf{Link Purpose (RQ3)}
Table~\ref{tab:purpose} shows the results of link purposes in commit messages and source code comments.
Providing metadata purpose has been dramatically reduced to 3--22\% in commit messages from 11--75\% in source code comments, as commit messages focus on the context of changes, rather than licenses or author information.} 

\fix{On the contrary, providing context purpose has been indirectly increased to 7--71\% in commit messages from 5--21\% in source code comments (especially for domains that are rarely linked). On the one hand, four link purposes (\texttt{commented-out source code}, \texttt{link-only}, \texttt{self-admitted technical debt}, and \texttt{@see}) become rare or even do not appear in commit messages. On the other hand, we also observe two new purposes, such as \texttt{version control sync} and \texttt{related issue}.}

\fix{These comparison results indicate that developers tend to reference short-lived link targets to keep files in sync. Since the main purpose of the commit message is to identify changes in this commit, developers tend to reference diverse link targets with the purpose of providing additional information related to the commit.}


\fix{\textbf{Link Target Evolution (RQ5)}
We can compare the results of the Stack Overflow case study to links in source code comments~\citep{hata20199}. In Figure~\ref{fig:so-case}, the median values of each change in commit messages share the same trend in source code comments, i.e., Comments \textgreater~Text Edits \textgreater~Answers \textgreater~Title Edits. However, only 91 (24\%) of the 372 Stack Overflow threads in source code comments did not undergo any change. \fix{These comparison results indicate that} Stack Overflow links referenced in commit messages tend to be less prone to change than those referenced in source code comments.}


\fix{\textbf{Link Decay (RQ6)}
When comparing this result to links in source code comments~\citep{hata20199},
the link decay in commit messages is heavily worse than in source code comments from this quantitative analysis, i.e., 70\% of links in commit messages and 18.8\% of links in source code comments are inaccessible. These comparison results indicate that commit messages often contain links to old systems or services and are in danger of information loss due to broken links.}

As we have \fix{discussed above}, the behaviors that developers follow to reference external resources in the two different software artifacts are different. Source code comments and commit messages are two important software artifacts in the development process that developers can use to share knowledge related to code or commits with their communities. It is important to understand the role of links in these two software artifacts, since compared to natural language, links contain richer information that could be useful in understanding code or commits.

In our paper, we find that developers tend to reference more links from fewer domains (RQ1) in commit messages; however, these links are more likely to be short-lived and related to the commit process with the purpose of keeping files in sync (e.g., bug report, pull request, and patch) (RQs1--3). To include complete information to describe the commit to the extent possible, developers reference diverse links (e.g., software, organization, application, API documentation, and forum thread) (RQ3). The link decay in commit messages is substantially worse than in source comments, especially for links to SVN repositories (RQ6). This issue could be a major cause of knowledge loss related to commit messages. In a case study of links pointing to Stack Overflow, we find that Stack Overflow links referenced in commit messages tend to be less prone to change than similar links in source code comments (RQ5).

In summary, we can see that links in source code comments are complementary to the source code, they provide relatively longer-lived information to help developers easily understand the metadata of code (i.e., license or software homepage as basic information for this code). In terms of the commit messages, links are more likely to be used as additional information related to changes or as a sign for syncing files to explain changes.

\subsection{Recommendations}
Based on our findings, we make the following recommendations for developers and researchers. First,
we recommend to developers:
\begin{itemize}
\item \textbf{\textsf{Pay special attention to maintaining commit related links}}. We find that developers tend to use links to reference software artifacts that are short-lived and related to the commit process (e.g., bug report, pull request, and patch). Developers should pay special attention to maintaining these links, since these software artifacts are important for understanding a commit. Moreover, these short-lived software artifacts can cause knowledge loss in the community.

\item \textbf{\textsf{Reference permanent links in commit messages}}. We find that \minor{inaccessible} links are prevalent in commit messages (34--43\% of links in our sample and 70\% of the distinct links in our data set). Commit messages are a critical means of communication between developers and reviewers, and links in commit messages are special containers for providing external and internal knowledge. Building stable traceability links between diverse software artifacts (e.g., bug \fix{reports}, \fix{patchs}, and API documentation) and commits could support knowledge sharing in the community.

\item \textbf{\textsf{Fix \minor{inaccessible} links when generating release notes from commit messages}}. We find that link decay is a common issue in commit messages. Commit messages can be used to generate release notes. \fix{\cite{9796331} analyzed 1,731 GitHub issues that are related to release notes, they found that 20.81\% of these issue reports complained about wrong or broken links in the release note.} Thus, to avoid suffering from the link decay issue in release notes, \fix{developers} should fix or omit these \minor{inaccessible} links from the release notes generation.

\end{itemize}



Second, we recommend to researchers: 
\begin{itemize}


\item \textbf{\textsf{Further studies of traceability links between other software artifacts and commits}}. We find various link targets exist in commit messages, e.g., patch, bug report, and tutorial or article. Bug reports have been studied, for example, \cite{sun2017frlink} recover missing issue-commit links by revisiting file relevance. We suggest that researchers should investigate traceability links between other software artifacts and commits.

\item \textbf{\textsf{\fix{Estimate the impact of the link decay in commit messages.}}} \fix{In \textbf{RQ2} and \textbf{RQ3}, we find that role of the \minor{inaccessible} links is associated with the purpose of keeping files in sync. We argue that link decay is indeed a problem, as it renders the information resource useless (i.e., a reviewer or user cannot access the link that holds supplementary information). Since this breakdown in information will result in a breakdown in communication and knowledge acquisition in the project, our plan is to detect cases where a \minor{inaccessible} link generates a discussion among interested parties. Depending on the situation, these removals of informal may cause delays in review time, increased discussion, and overall acceptance of code commits. Analyzing such characteristics, for example, could shed light on the negative impact of \minor{inaccessible} links on software development.}

\item \textbf{\textsf{Tools to support common commit message templates}}. We find that links in commit messages point to a smaller range of different domains compared to links in source code comments, suggesting that there are common use cases for links in commit messages. A promising research direction could be the creation of templates for commit messages to increase the probability that developers remember to add links where relevant, for example, following the work on using stereotypes to characterize commits~\citep{dragan2011using}.


\end{itemize}

\section{Threats to Validity}
\label{sec:tv}
In this section, we discuss the threats to the validity of our study.

\fix{\textbf{Construct Validity.} To compare the role of links between source code comments and commit messages, we used the same stratified sample of repositories as the previous study~\citep{hata20199}. However, the use of the same list of repositories may introduce an inconsistency in the comparison due to the presence of private or unavailable repositories. To keep the consistent comparison between links in source code comments and commit messages, we only consider http(s) links as URLs in commit messages by the regular expression. For work on other kinds of links in the context of software development, we refer readers to ~\cite{schermann2015discovering} who studied the interlinking characteristics in commits and issues. Future work is needed to gain a global picture of links in commit messages \minor{for different URL protocols, e.g., FTP and SSH}.}

\textbf{Content Validity.} In our study, we manually classified a sample of 1,145 links in commit messages, which carries the risk of undiscovered link targets and link purposes in commit messages. \fix{Since we are unable to reliably infer \minor{the type and} the intended purpose of links for which the target is no longer accessible, we decided to code such links consistently as \minor{\texttt{unknown (404)} similar to previous study~\citep{hata20199}}. Trying to guess the type and intended purpose for a subset of these links \minor{(e.g., \url{https://svn.apache.org/repos/asf/subversion/trunk@851235} could be coded as a \texttt{patch})} would likely have had an impact on the results reported for link types and purposes \minor{since the target content is not always obvious from the link alone and since not all links were archived in a consistent manner in Wayback Machine or similar services.}}
\fix{Finally, as part of the qualitative analysis, we defined three strata of samples generated visually from the distribution of the data. Although subjective, we are confident in this sampling, especially since this data does not follow a normal Gaussian distribution.}

\textbf{Internal Validity.} To answer RQs2--4, we conducted qualitative studies of a sample of all links in our data set. These qualitative studies are manual analyses that were conducted according to our coding guides. These codes may be inadequate due to the subjective nature of understanding the coding guides. To \fix{mitigate} this threat, we require \fix{Fleiss's k}appa agreements~\citep{fleiss1971measuring} of at least ``Substantial agreement'' for the understanding of the coding guides in RQs2--3, following previous work~\citep{wang2021understanding, xiao2021characterizing}.
In RQ4, we define the coding guides as rules that automatically classify links. The rules are dependent on the link target domains and the coding results of RQ2.  

\textbf{External Validity} Although we analyzed a large number of commit messages from GitHub repositories, our
results may not generalize to industry or other open-source artifacts, in general. Some open-source repositories are hosted outside of GitHub, e.g., on GitLab or private servers.

\section{Related Work}
\label{sec:rw}
In this section, we discuss existing work related to commit messages, knowledge sharing, and link sharing.

\subsection{Commit Messages}
In modern software development, developers submit commits to version control systems to integrate new features or fix bugs.
Commit messages are required to document or summarize such changes.
While the size and complexity of software systems grow with an increasing number of commits, those commit messages become critical to understanding code changes, especially if issues occur. 

In the study of~\cite{alali2008s}, they found that commit messages are coupled with three size-based characteristics of commits (number of files, lines, and hunks).
As external documentation of source code changes, commit messages play a critical role in open-source projects.
Open-source projects often have rules about commit messages to ensure governance, but as a community based on voluntary contributions, it is difficult to enforce such rules~\citep{o2007emergence}. Therefore,~\cite{dyer2013boa} observed that there have been around 14\% of commit messages in more than 23,000 open-source SourceForge Java projects that are completely empty. \cite{santos2016judging} used the n-gram cross entropy of text in commit messages to successfully identify commits that were likely to make a build fail. Commit messages are also used as a link between issue reports and commits~\citep{xie2019deeplink}, recommend refactoring opportunities~\citep{rebai2020recommending}, detect and classify refactoring descriptions~\citep{krasniqi2020enhancing}, identify security issues~\citep{zhou2017automated}, and identify whether this commit can be skipped by continuous integration~\citep{abdalkareem2020machine}. 
However, due to the voluntary contribution of open-source projects and time pressures of developers~\citep{o2007emergence, maalej2009work, murphy2009attacking, d2010commit}, commit messages can be non-informative, lack information, meaningless, or completely empty~\citep{maalej2010can, tian2022makes, dyer2013boa, liu2020atom}. Therefore, many researchers have proposed commit message generation tools to facilitate \fix{the} understanding of commit changes. For example,~\cite{liu2020atom} described ATOM, a tool to encode abstract syntax tree paths of diffs to generate commit messages. Similarly,~\cite{huang2020learning} presented ChangeDoc, an approach to generate commit messages from existing messages. \fix{In this study, we recommend that developers include permanent links in their commit messages to prevent knowledge loss during the commit process. However, it is possible that developers may not always follow this recommendation. Therefore, we suggest implementing an approach to automatically generate commit messages and archive links for changes, which would facilitate the preservation of knowledge during the commit process.   }

Moreover, commit messages are critical for knowledge transfer during project maintenance. \cite{fu2015automated} presented a semi-supervised Latent Dirichlet Allocation based approach to automatically classify change messages (i.e., Corrective, Perfective, and Adaptive). In validation surveys, they found that this approach can be applied to cross-projects; automatic classification results can reach around 70\% in agreements with developers.
\cite{sarwar2020multi} also presented an approach using transfer learning to classify commit messages into the same categories. Commit messages help developers understand changes and their underlying rationale with less effort, process influences, and less bias between communication.~\cite{mockus2000identifying} identified four types of changes based on commit messages: adding new functionality, repairing faults, restructuring the code to accommodate future changes, and code inspection rework. The messages help to reduce efforts on other software development tasks: generate release notes~\citep{moreno2014automatic}. \fix{In this study, we did not examine the relationship between types of links and types of changes. However, we plan to address this aspect in future research. }

Unlike the aforementioned studies, our study investigates links as one representation of knowledge in commit messages. Investigating links in commit messages enables us to understand the role they play as well as potential issues with links in commit messages.
In our notion, links form an important communication channel between the initial code authors and later reviewers and maintainers.


\subsection{Knowledge Sharing}

Knowledge sharing is a central aspect of software development.
Knowledge is transferred or exchanged among people and communities on online platforms, for example, Wikipedia~\citep{kittur2010beyond,forte2012coordination,nagar2012you} and GitHub~\citep{dabbish2012social}.
\cite{dabbish2012social} interviewed light and heavy GitHub users, and showed that developers made a variety of social inferences about other developers and
projects to collaborate, learn and manage projects using the networked activity information.
This study also suggested that transparency of such the networked activity information 
can support knowledge sharing, new ideas, and community.
\cite{wattanakriengkrai2022github} investigated the referencing of academic papers in README files of GitHub repositories. They found that this knowledge sharing is rarely occurring (0.4\%), however, the majority of these referenced academic papers are open access (98.5\%). \fix{We also found a few links referenced to academic papers in commit messages, which indicate such knowledge sharing is also rarely occurring in commit messages. }

In addition to GitHub, \cite{aniche2018modern} surveyed the Reddit programming subreddit and analyzed Reddit and Hacker News posts. They found that development posts are the highly occurring theme, and these posts are mostly related to programming or markup languages. 
Developers on the r/programming subreddit aim at learning, but on Hacker News, they focus more on publicizing news. They also suggested researchers expand Reddit and Hacker News as sources for developer knowledge. Furthermore, other research focuses on knowledge sharing on Q\&A sites~\citep{movshovitz2013analysis, vasilescu2014social, baltes2022contextual}, between Q\&A sites and GitHub~\citep{vasilescu2013stackoverflow}, and GitHub discussions~\citep{hata2022github}. \fix{Inspired by these works, we expand links in commit messages as a source for developer knowledge.}


Although plenty of studies widely investigate knowledge sharing in software engineering, there is no study that focuses on
links as special containers that provide additional knowledge for developers in commit messages. In this study, we focus on the different link targets and purposes that they served on commit messages, rather than natural language, to study knowledge sharing.

\subsection{Link Sharing}
As one convenient way of knowledge sharing, link sharing has been widely adopted and explored in developer communities (i.e., Q\&A sites, GitHub, code review). \cite{gomez2013study} found that a significant proportion of links shared on Stack Overflow, in particular, were used to transfer knowledge about software development innovations such as libraries and tools. \cite{ye2017structure} also analyzed link sharing activities in Stack Overflow to study the structural and dynamic properties of the emergent knowledge network in Stack Overflow. They discovered that developers share links for diverse purposes, e.g., reference information for problem solving is the most occurring purpose. In addition to that, external links are investigated on aspects of broken links~\citep{liu2021broken} and repeatedly referenced links~\cite{liu2022exploratory} in Stack Overflow posts.

A previous study~\citep{hata20199} analyzed 9.6 million links that exist in source code comments. They explored that link sharing is common in source code comments, more than 80\% of the repositories contained at least one link. They also identified the kinds of link targets (i.e., licenses, software homepages, and specifications) and link sharing purposes (providing metadata or attribution).  \fix{\cite{zampetti2017developers} investigated
to what extent and for which purpose developers refer to external online resources in pull requests. The findings of their investigation suggest that developers frequently consult external resources for the purpose of acquiring new knowledge or resolving specific issues.}~\cite{aghajani2019software} presented an empirical study which shows ``Outdated/Obsolete references'' are one of the issues in software documentation. 
Furthermore, \cite{baltes2019usage} found that 40\% of their survey participants added a source code comment in GitHub projects with a Stack Overflow link to the corresponding question or answer. In addition to the Stack Overflow link to a question or answer, issue report links were studied. \cite{zhang2018within} showed that developers tend to link more cross-project or cross-ecosystem issues over time. \cite{zhang2020ilinker} proposed an approach called iLinker for issue knowledge acquisition in GitHub projects, it can improve the development efficiency of GitHub projects. \fix{Similar to previous work, we also analyzed links in GitHub artifacts. We found that link decay is a frequent phenomenon in commit messages. Moreover, developers reference links for purpose of providing context. }

\cite{rath2018traceability} addressed missing links between commits and issues, and proposed an approach to generate the missing links by a combination of process and text-related features. The practice of link sharing was also studied in the context of code review. \cite{wang2021understanding} performed a mixed-method approach to highlight the role that shared links play in the review discussion. Their results show that the link is served as an important resource to fulfill various information needs for patch authors and review teams. \fix{\cite{liu2022method} proposed an approach to identify references between projects by extracting links from pull requests, issues, and commits. These links were obtained by patterns of URL, ID of issue, pull request or commit.} \fix{Unlike these
works in the code review process, we focus on the comparison between the roles of links
in commit messages and source code comments to obtain an understanding of informal and initial
knowledge sharing in the software development process.}

Inspired by these past studies of link sharing, we conduct the first study on links in commit messages. Similar to prior work, we investigate the prevalence, targets, purposes of links and the phenomenon of links (i.e., repeated link reference, link evolution, and link decay) in commit messages.

\section{Conclusion}
\label{sec:cs}
In this paper, we conducted: (i) a quantitative study of 18 million links from commit messages in 23,110 Git repositories to investigate the prevalence of links in commit messages; (ii) qualitative studies of a stratified sample of 1,145 links to determine the kinds of link targets, the purposes of referencing these links and their repeated links; (iii) a quantitative study to investigate the evolution of link targets; and (iv) a quantitative study to investigate how the link decay issue is common in commit messages and investigate which domains frequently affect link decay.

We observed that (i) links are frequently occurring in commit messages, accounting for at least 83\% of GitHub repositories in our study; (ii) 34--43\% of links in our sample are unavailable, other than that, links to patch are common in commit messages; (iii) the purpose of adding additional information is the most prevalent in commit messages; (iv) repeated links are rarely occurring in commit messages, accounting for four percent of our sample; (v) 14\% of the links are prone to evolve to become unavailable over time; and (vi) link decay is a common issue in commit messages, around 70\% of the distinct links in our study are \minor{inaccessible}. We foresee many promising avenues for future
work, such as tool support to fix \minor{inaccessible} links, expanding our
coded corpus to other software artifacts, and
further studies of commit messages.

\section*{}

\textbf{Acknowledgements} This work was inspired by the International Workshop series on Dynamic Software
Documentation, held at McGill’s Bellairs Research Institute, and was supported by JSPS KAKENHI Grant Numbers JP18H04094, JP20K19774, JP20H05706, and JP22K11970 and JST PRESTO Grant Number JPMJPR22P6.
\\
\\
\noindent \textbf{Data Availability} The datasets generated during and/or analysed during the current study are available in
the Zenodo repository, \url{https://doi.org/10.5281/zenodo.7536500}.
\\
\\
%
\section*{Declarations}
\textbf{Conflict of Interests} The authors declare that Sebastian Baltes, Hideaki
Hata, Christoph Treude, and Raula Gaikovina Kula are members of the EMSE Editorial Board.
All co-authors have seen and agree with the contents of the manuscript and there is no financial interest to report.

\bibliographystyle{spbasic}      

\bibliography{main}   

\end{document}